\documentclass[prl,twocolumn,showpacs,superscriptaddress,floatfix]{revtex4}

\usepackage{amsmath}			
\usepackage{amssymb}			
\usepackage{graphicx}			
\usepackage{epic}

\begin{document}

\title{Structure of fermion nodes and nodal cells}

\author{Lubos \surname{Mitas}}
\affiliation{Department of Physics and CHiPS, 
North Carolina State University, Raleigh,
NC 27695}

\date{\today}

%
\begin{abstract}
We study nodes of fermionic ground state 
wave functions.
For $2D$ and higher we prove that spin-polarized,
noninteracting fermions in 
a harmonic well have two nodal cells for arbitrary system size.
The result extends to other noninteracting/mean-field models
such as fermions on a sphere, 
in a periodic box or in Hartree-Fock atomic states. 
Spin-unpolarized noninteracting states have multiple nodal cells,
however, interactions and many-body correlations generally relax the
multiple cells to the minimal number of two. 
With some conditions, this
is proved for interacting
$2D$ and higher dimensions harmonic fermion systems of arbitrary
size using the Bardeen-Cooper-Schrieffer variational
wave function. 
Implications and extent of these results are briefly discussed.
\end{abstract}


\maketitle

%
A fermion node is a subspace of fermion configurations for which 
a real wave function describing the fermionic system 
vanishes due to the antisymmetry. 
In general, for $N$ spin-polarized fermions in $d$ dimensions
the fermion node is a $(dN-1)$-dimensional hypersurface given by 
an implicit
equation $\Psi(R)=0$, where 
$\Psi$ is the wave function with fermion 
coordinates $R=({\bf r}_1, ..., {\bf r}_N)$.
The location of the nodal manifold is
of key importance for quantum Monte Carlo (QMC) methods 
\cite{qmchistory, hammond, qmcrev,jbanderson} since 
the exact node enables us to 
solve the stationary Schr\"odinger equation
with computer time scaling as
a low-order polynomial in $N$. 
Remarkably, even rather crude 
Hartree-Fock (HF) or post-HF wave function nodes, 
routinely used in the fixed-node approximation QMC methods, 
provide $\approx$ 95\% of the correlation energy in a variety
of systems with hundreds
of valence electrons \cite{qmchistory,hammond,qmcrev,jbanderson}.
However, to reach beyond this level of accuracy 
has proved to be challenging
because of limited
understanding of fermion nodes 
 \cite{jbanderson,davidnode,lesternode,sorella,foulkes,dariobe}. 



The exact nodes for interacting systems are known only for a very 
few two-electron triplet atomic states
\cite{dariohe,davidnode} and, very recently, 
the exact node of three-electron 
$^4S(p^3)$ state has been discovered
\cite{bajdich}. It turned out
that the nodes of these high-symmetry states 
have rather simple topologies and divide the configuration
space into two compact 
nodal cells in which the wave function is positive 
or negative ("plus" or "minus" cell).
The two nodal cells were found also in      
noninteracting spin-polarized $2D$ and $3D$ fermions
with up to 200 particles using a numerical proof
\cite{davidnode}. 
The same noninteracting but spin-unpolarized systems trivially have
four nodal cells since the wave function is a product 
of spin-up and spin-down determinants. 
Interestingly, analysis of interacting unpolarized few-particle 
systems \cite{dariobe,cmt28,lesternode} has revealed
that the electron correlation can change the
node topologies and number of nodal cells. 
For example, the correlated wave functions
of the Be atom or N$_2$ molecule exhibit 
two nodal cells while the corresponding
Hartree-Fock ones have four.
It was therefore conjectured that
the bisection of the configuration space into 
the two nodal cells might be a generic property of 
fermionic ground states 
(with exceptions discussed later).
In this Letter we prove that for $2D$ spin-polarized 
noninteracting harmonic fermions 
the ground state node
divides the configuration space into the minimal number of two nodal
cells for any system size. 
The proof method extends to higher dimensions
and carries over to other models such
as fermions on a sphere, in a periodic box, for atomic states, etc. We
show that the same holds, in general, for
interacting systems using correlated Bardeen-Cooper-Schrieffer
wave functions and we briefly discuss the implications.

We recall two basic properties of fermion nodes 
derived by Ceperley \cite{davidnode}. 
a) Nondegenerate ground state wave functions satisfy the so-called 
{\it tiling property} which states 
 that by applying all possible particle permutations to an arbitrary
nodal cell one covers the entire configuration
space. Note that this does not specify the number of nodal cells.
b) Consider three particles $i,j,k$ in a spin-polarized system
with  wave function $\Psi(R)$.
We call the particles $i,j,k$ connected,
if there exists a triple exchange $ijk \to jki$
path that does not cross the node, ie,  $|\Psi(R)|>0$ along the exchange path. 
More connected particles, such as the following six ones, can form a cluster:
\setlength{\unitlength}{0.5mm}
\begin{picture}(12,7.5)(6,4.5)
\put(2,5){$\bullet$}
\put(6,2){$\bullet$}
\put(6,8){$\bullet$}
\put(10,5){$\bullet$}
\put(14,2){$\bullet$}
\put(14,8){$\bullet$}
\drawline(3.5,7)(7.7,4)(7.7,10)(3.5,7)
\drawline(7.7,10)(15.8,4)
\drawline(7.7,4)(15.8,10)(15.8,4)
\end{picture}.
If there exists a point $R_t$ such that
triple exchanges {\it connect all the particles into
a single cluster} then  $\Psi(R)$
has only {\it two nodal cells}.
The tiling property implies that once the particles
are connected for $R_t$  
the same applies to any point in the cell as 
further explained in Ref. \cite{davidnode}.   

We illustrate the properties of nodes in $1D$ using 
spin-polarized fermions 
in a $1D$ harmonic oscillator well.
The wave function is a Slater determinant  
$\Psi(1,...,N) = {\rm det}\{\phi_k({\bf r}_i)\}=
A\prod_i e^{-x_i^2/2}{\rm det}\{1,2x, ...,H_{N-1}(x)\}$ 
where $H_n(x)$ is a Hermite polynomial of degree $n$ and $A$
is the normalization.
We omit the prefactors and transform 
the Slater matrix 
to monomials so that 
the wave function is given by the Vandermonde determinant
\begin{equation}
\Psi_{1D}(1, ...,N)={\rm det} \{1,x,x^2, ..., x^{N-1}\}
= \prod_{i<j}(x_j-x_i)
\label{eq:oned}
\end{equation}
The node is encountered whenever two fermions pass through
each other and the wave function has
 $N!$ nodal cells since any permutation requires at least
one node-crossing. In general, the derived node is exact for other
$1D$ models including systems with interactions.

Now consider spin-polarized fermions in a $2D$ harmonic well.
The one-particle states are
simply $\phi_{nm}=C_{nm}H_{n}(x)H_{m}(y),$ $n,m=0,1, ... $
where $C_{nm}$ includes the gaussian and normalization which are
absorbed into a common prefactor and omitted. 
The Slater matrix elements can be rearranged 
to monomials and we write 
\begin{equation}
\Psi_{2D}(1,...,N)={\rm det}\{1,x,y,..., x^ny^m, ...\}
\label{eq:2dmono}
\end{equation}
The closed-shell states and the system size are labeled by 
$M=1,2, ...$ 
where $n+m\le M$ with the number of fermions given by $N=(M+1)(M+2)/2$.

Using induction we prove that the wave function 
in Eq. \ref{eq:2dmono}
has only two nodal cells for any $M>0$. 
This is indeed true 
 for $M=1$ with three-particle wave function 
$\Psi_{2D}(1,2,3)={\rm det}\{1,x,y\}$. 
In order to show this, it is convenient to
extend the particle $2D$ coordinates 
by a "dummy" third dimension as ${\bf r}_i=(x_i,y_i,0)$.
Then $\Psi_{2D}(1,2,3)={\bf z}_0\cdot ({\bf r}_{21}\times
{\bf r}_{31})$ where ${\bf z}_0$ is the unit vector in 
the third dimension
and ${\bf r}_{ij}={\bf r}_i-{\bf r}_j$. Clearly, there are only two nodal
cells since the set of vectors ${\bf z}_0, {\bf r}_{21}, {\bf r}_{31}$ is
either left- or right-handed, and the node is encountered whenever
the three particles are
collinear, ie, ${\bf r}_{21}\times{\bf r}_{31}=0$. 
Also, the particles are connected 
by triple exchanges without node-crossing (eg, rotate an equilateral triangle). 

\begin{figure}[ht] 
\centering
{\resizebox{2.5in}{!}{\includegraphics{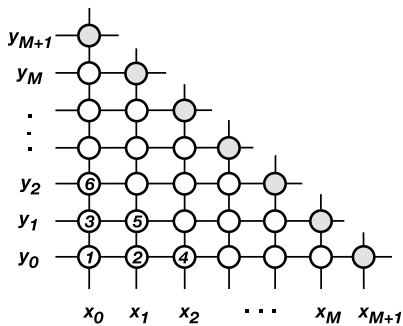}}}
\caption{Particles arranged  
in a Pascal-like triangle pattern. The circles denote the particle positions 
and inside the circles are particle labels. }
\label{fig:pascal}
\end{figure}

We now consider a general system with $M>1$ and arrange
the particles into a Pascal-like triangle pattern on a rectangular
mesh as shown in 
Fig.\ref{fig:pascal}. 
For this arrangement  
the determinant can be explicitly evaluated for any $M$
by subsequent factorization of lines of particles \cite{lagr}.
Due to the space constraints and to simplify the notation
we illustrate the factorization on a few-particle example;
the generalization to an arbitrary size is straightforward.
For example, for $M=2$ the wave function reads
\begin{equation}
\Psi_{2D}(1,...,6)={\rm det}\{1,x,y,x^2,xy,y^2\}=\nonumber
\end{equation}
\begin{equation}
=\left| \begin{array}{cccccc}
1   &  1        &  1   & 1 & 1 & 1   \\
x_0 &  x_1      &  x_0 & x_2 & x_1 & x_0 \\
y_0 &  y_0      &  y_1 & y_0 & y_1 & y_2 \\
x_0^2 &  x_1^2    &  x_0^2 & x_2^2 & x_1^2 & x_0^2 \\
x_0y_0 & x_1y_0 &  x_0y_1 & x_2y_0 & x_1y_1 & x_0y_2 \\
y_0^2 &  y_0^2  &  y_1^2 & y_0^2 & y_1^2 & y_2^2 \\
\end{array} \right| 
\end{equation}
where the particle positions are given in Fig.\ref{fig:pascal} (our point $R_t$). 
Clearly, all the last-row elements containing
$y_0$ can be eliminated
by adding a multiple of
the third row.  In a similar way we eliminate
all the matrix elements containing $y_0$, and
the determinant factorizes as
\begin{equation}
\Psi_{2D}(1,...,6)= 
\prod_{i=1}^2 (y_i-y_0)^{n_i}\Psi_{1D}(1,2,4)\Psi_{2D}(3,5,6)
\end{equation}
where $n_i$ is the number of particles on line $y=y_i$. 
Note that one of the factors is
the $2D$ wave function for the system  
with $M$ reduced by one.
The same factorization structure is obtained for an arbitrary $M$ 
and, obviously, it can be applied recursively. The wave function   
for a general size $M$ is then given by
\begin{equation}
\Psi_{2D}(1,...,N)=\pm\prod_{i>0}^M (y_i-y_0)^{n_i}
\Psi_{1D}(I_0)\Psi_{2D}(1,...,N/I_0)
\nonumber
\end{equation}
\begin{equation}=\pm
\prod_{l=0}^{M-1} \left[\,\prod_{i>l}^M  
(y_i-y_l)^{n_i}\Psi_{1D}(I_l)\right]
\label{eq:recur}
\end{equation}
where $I_l=i^{(l)}_1, ..., i^{(l)}_{M+1-l}$ 
denotes the labels of particles 
lying on the line $l$ while $(1,...,N/I_0)$
means that the labels in $I_0$ are omitted from $1,...,N$.
The sign depends on the number of row exchanges 
and on the actual ordering of particles. Note the translation
invariance of Eq. \ref{eq:recur}.

For the induction step we assume that triple
exchanges connect all the particles in the system of size $M$.
Let us show that the same is true for the system of size $M+1$
(see Fig.\ref{fig:pascal}). 
We first factorize out the line $y=y_0$ as given 
by Eq. \ref{eq:recur}. One of the factors is the
$2D$ wave function for the system of size $M$, which contains all additional 
particles except the one with coordinates $[x_{M+1},y_0]$. 
By the original assumption,
such a system has all the particles connected.
 If instead of the horizontal line we factorize out
 the vertical line ($x=x_0$) we see
 that the particle at
$[x_{M+1},y_0]$  is also connected, 
thus concluding the proof. 

The proof can be generalized in several ways. 
The same arguments apply to
$3D$ since particles can be arranged into
an appropriate $3D$ pattern and the wave function evaluated 
by recursive factorization of {\it
planes and lines}. In fact, the result holds and the 
two nodal cell property is correct for  arbitrary dimension $d>1$! 
The proof applies also
to other noninteracting or mean-field models with
polynomial entries in the Slater matrix such as
fermions on a sphere or in a periodic box   
with details given elsewhere \cite{mitaslong}.
The proof can be modified at least for some open-shell 
states as well while taking 
into account node ambiguities if there are degeneracies 
\cite{davidnode, foulkes}.

The proof method can be further combined with symmetries
for cases when a complete factorization is not obvious, such
as for multi-shell atomic states. 
For example, spin-polarized ground states 
$1s2s2p^n, n=1-3$ (and beyond) also have
two nodal regions. We show this for the
$^6S(1s2s3p^3)$ state in the noninteracting and HF limits. 
The wave function is written as 
 $\Psi_{at}(1,...,5)={\rm det}\{\rho_{1s}^{*}(r),\rho_{2s}^{*}(r),x,y,z\}$
where $\rho_{1s}^{*}(r)=\rho_{1s}(r)/\rho_{2p}(r)$ and 
 $\rho_{2s}^{*}(r)=\rho_{2s}(r)/\rho_{2p}(r)$ since nonnegative
$\rho_{2p}(r)$ is factorized out. The dimensionless coordinates 
are rescaled by the atomic number $Z$ and
the Bohr radius $a_0$ as ${\bf r}\leftarrow Z{\bf r}/a_0$.
Let us place particle 1 at the origin and particles 2-5 on the surface
of a sphere with the radius $\eta_0$ equal to the radial node of 
$\rho_{2s}(r)$ orbital,
ie, $\rho_{2s}(\eta_0)=0$. For such configurations we obtain 
$$
\Psi_{at}(1,...,5)=
\rho_{1s}^{*}(\eta_0)\rho_{2s}^{*}(0)\, {\bf r}_{32} \cdot
({\bf r}_{42} \times {\bf r}_{52})
$$
so that any three-particle exchange from 2,3,4,5 easily avoids 
node-crossing by appropriate positioning and rotations on the sphere.
The particle 1 is connected by
the exchange $123\to 312$ parametrized 
as ${\bf r}_1(t)=\eta_0[t,0,0]$; ${\bf r}_2(t)=\eta_0[c(t),
s(t), 0]$ and ${\bf r}_3(t)=\eta_0[0,1-t,0]$ where $t=0$ ($t=1$) 
corresponds to the beginning (end) point of the exchange path while $c(t)=\cos(\pi t/2)$ and $s(t)=\sin(\pi t/2)$. Setting
 ${\bf r}_4=[0,0,\eta_0]$ 
and  ${\bf r}_5=[0,0,-\eta_0]$ we find that $\Psi_{at}$ is proportional to
$$
\rho_{2s}^{*}(t\eta_0)c(t)(1-t)+ \rho_{2s}^{*}[(1-t)\eta_0]s(t)t>0
$$ 
The inequality holds for the whole path
$0\leq t\leq 1$ since  $\rho_{2s}^{*}(t\eta_0)>0$ for  $0\leq t <1$ 
for both noninteracting and HF cases. 
The proof can be further extended to more shells such 
as $1s2s2p^33s3p^3$ and $1s2s2p^33s3p^33d^5$ \cite{mitaslong}.

{\it Spin-unpolarized systems.} Due to the 
product of spin-up and -down determinants,
the number of nodal cells in noninteracting
unpolarized systems is twice
the number of cells of the half-filled spin-polarized counterparts. 
The proof for atomic states above
then implies
that the HF wave functions for atoms with $Z>3$ up to 
the third-row elements have
four nodal cells.


{\it Interactions in spin-polarized systems.}
In general, the shape and topology of the 
nodal manifold is
influenced by interactions
and many-body  effects.
Consider the lowest atomic quartet of $S$ symmetry
and even parity $^4 S^e (1s2s3s)$
(clearly not the lowest quartet, which is odd $^4P^o(1s2s2p)$). 
The noninteracting wave function is
given by $\Psi_{at}(1,2,3)={\rm det } \{ \rho_{1s}(r), \rho_{2s}(r), \rho_{3s}(r)\}$ and has
{\it six} nodal cells.  Since $\Psi_{at}(1,2,3)$ depends only on  
distances it is quasi-$1D$ and the nodes 
are the same as the ones given by
Eq. \ref{eq:oned}. 
For the interacting system 
the correlation is included by adding to the wave function
the lowest (and dominant) double excitation 
$2s3s \to 2p_x3p_x + 2p_y3p_y + 2p_z3p_z $ with a weight $w$.
The correlated
 wave function then allows for exchanges without node-crossing.
Define ${\bf r}_a(t)=[0,c(t),s(t)]$,
${\bf r}_b(t)=[c(t),s(t),0]$,
 ${\bf r}_{ab}(t)=[1-g(t)]{\bf r}_b(t)-g(t){\bf r}_a(t)$, 
${\bf r}_c(t)={\bf r}_{ab}(t)/|{\bf r}_{ab}(t)|$,
and
${\bf r}_d(t)={\bf r}_c(t)\times {\bf r}_a(t)$, 
where $g(t)=3t(1-t)$.
The exchange path $123\to 231$ is then ${\bf r}_{3}(t)=r_3(t){\bf r}_a(t)$,
${\bf r}_{2}(t)=r_2(t){\bf r}_c(t)$ and
${\bf r}_1(t)=r_1(t){\bf r}_{d}(t)/|{\bf r}_{d}(t)|$.
The radial parts are given by 
$r_1(t)=\eta_m+q(2t-1)$, $r_2(t)=\eta_m+q(1-t)$  and 
$r_3(t)=\eta_m-qt$. $\eta_m$ is the mean value of the  
radial node of $\rho_{2s}$ and the first radial node of $\rho_{3s}$ 
orbitals while  $0<q< a_0|w|$ (for the Coulomb e-e interaction
$w$ is  $ \approx - 0.05 $).
The path is orchestrated so that in the region where
the noninteracting component vanishes the correlation dominates 
and the particles become connected.
This illustrates two points:
imposing symmetries
at the mean-field level can generate multiple cells and, in general, 
for $d>1$ the
interactions lift this "nodal cell degeneracy" and relax
multiple cells to the minimal two.

{\it Interactions in spin-unpolarized systems.} 
The change from four to two nodal cells due to interactions 
has been demonstrated
for the first time
on the Be atom \cite{dariobe} using the two-configuration correlated wave
function and  the connected cluster construction 
adapted to spin-unpolarized systems.
Consider a simultaneous exchange of one (or more) pair(s) 
of  spin-up particles and one (or more) pair(s) of spin-down particles.
For noninteracting wave functions such simultaneous pair exchanges
imply that the node will be crossed once or multiple times.
If there exists a point $R_f$ such that during the simultaneous 
spin-up and -down pair exchanges the inequality
   $|\Psi|>0$ holds along the whole path,
 then the wave function has only two nodal cells.

Consider a singlet state of $2N$ particles in a $2D$ harmonic well with 
particles interacting by pair potential. 
With some restrictions, we show that the correlation included in
the Bardeen-Cooper-Schrieffer (BCS) pairing
wave function \cite{sorella,carlson} given by
$\Psi_{BCS}(1,...,2N)={\rm det} [\Phi(i,j)]$ is enough to 
eliminate the noninteracting four nodal cells and fuse them 
into the minimal two. 
Here $\Phi(i,j)=\Phi(j,i)$ is a singlet pair orbital for
$i\negthinspace\negthinspace\uparrow $ and
$j\negthinspace\negthinspace\downarrow$ fermions and we decompose it
into noninteracting and correlated components
$\Phi(i,j)= \Phi_{0}(i,j) + \Phi_{corr}(i,j)$. Using one-particle orbitals
we can write 
$\Phi_{0}(i,j)=\sum_{n+m=0}^M\negthinspace \phi_{nm}(i)\phi_{nm}(j)$
while $ \Phi_{corr}(i,j)=\sum_{n+m>M}c_{nm}\phi_{nm}(i)\phi_{nm}(j)$
where $\{c_{nm}\}$ are variational parameters.

\begin{figure}[ht]
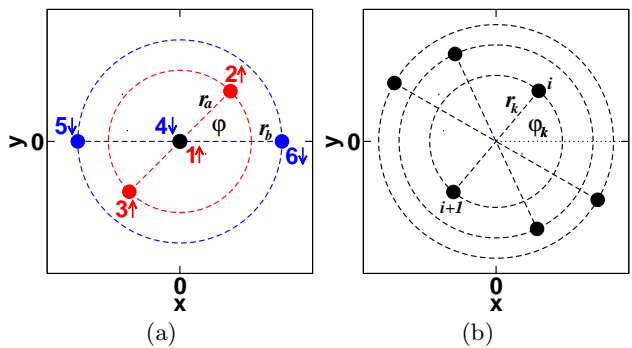

\centering
\begin{tabular}{cc}
\includegraphics[width=1.60in,clip]{figure2a.eps} &
\includegraphics[width=1.60in,clip]{figure2b.eps} \\
(a) & (b)
\end{tabular}
\caption{(a) Positions of six $2D$ harmonic fermions with two particles at
the origin and two pairs on circles with radii $r_a,r_b$.
(b) Positions of the spin-up particles for the $M=2,$ $2N=12$
particle singlet. The spin-down positions pattern is
similar except for the different radii and angles.}
\label{fig:polyex2}
\end{figure}

We illustrate this first on a six-particle singlet, $M=1,$ $2N=6$. 
In cylindric coordinates 
$(r,\varphi)$ we find $\Phi_{0}(i,j)=1+2r_ir_j\cos\varphi_{ij}$ 
where $\varphi_{ij}=\varphi_i-\varphi_j$,  
with the gaussians and normalization omitted.
$\Phi_{corr}(i,j)$ is constructed as a sum of 
orbitals from the next unoccupied shell $(n+m=2)$ and we find 
$\Phi_{corr}(i,j)=\alpha[2(r_ir_j\cos\varphi_{ij})^2 -r_i^2 -r_j^2],$
where a variational parameter $\alpha$ has been
included. We place the particles as in Fig.\ref{fig:polyex2}(a). 
For such a configuration $\Psi_{BCS}(1,...,6)=
8\alpha r_ar_b\cos\varphi[2(r_ar_b\cos\varphi)^2-r_a^2-r_b^2]$.  
The rotation of the system by $\pi$
exchanges the pair of particles for each spin, and 
 since the wave function is rotationally
invariant, it is enough to show that  
it is nonzero {\it for a single
point}, eg, $r_a=r_b=1, \varphi=\pi/4$. For $\alpha=0$ the wave function 
vanishes since the particles lie on the noninteracting node.
However, for {\it any} interaction which does not break the
rotation invariance and gives 
$\alpha \neq 0$,  the BCS
wave function has only two nodal cells.

Remarkably,
this can be generalized to an arbitrary size. Assume a closed-shell 
singlet with the total number 
of particles $2N=(M+1)(M+2)$ where
$N$  is even (for $N$ odd  the derivations are the same after placing
one particle of each spin to the origin, see Fig.\ref{fig:polyex2}(a)). 
We form $N/2$ pairs of particles in each spin subspace
so that, say, the pair $i\negthinspace\uparrow,
(i+1)\negthinspace\uparrow $ has coordinates 
given by $(r_i,\varphi_i)=(r_k,\varphi_k)$,
 $\,(r_{i+1},\varphi_{i+1})=(r_k,\varphi_k+\pi)$, 
see Fig. \ref{fig:polyex2}(b). Here $k=1,...,N/2$ labels the pairs 
in the spin-up channel; the spin-down particles are 
placed similarly and labeled by the pair index $l=N/2+1,...,N$. 
In this configuration the particles lie on the
noninteracting node since 
$ {\rm det}\{\Phi_0(i,j)\}=$ ${\rm det}\{\sum_{n+m\leq M} \phi_{nm}(i)
 \phi_{nm}(j)\}=$ ${\rm det}\{\phi_{nm}(i)\}
{\rm det}\{\phi_{nm}(j)\}$
and the rotation of the system by $\pi$ crosses the nodes 
in both spin channels
so that both Slater determinants vanish due to the rotation invariance.
Now, if all the $N$ pair distances and angles
$r_k,r_l,\varphi_k,\varphi_l$ are distinct, 
then each of the matrices $\{\phi_{nm}(i)\}, \{\phi_{nm}(j)\}$
has exactly one linearly dependent row, ie, their ranks are $N-1$. 
This can be verified directly
for small values of $M$ and then using induction for any $M$.  
Consequently, the matrix $\{\Phi_0(i,j)\}$ has linear dependence 
in one row and one column, ie, it has the rank of $N-1$ as well. In general,
adding virtual states through $\Phi_{corr}(i,j)$ provides 
independent rows/columns 
(eg, $M+1$ independent rows/columns from the first unoccupied shell)
which eliminate linear dependency so 
that ${\rm det}\{\Phi_0(i,j)+\Phi_{corr}(i,j)\}$ is nonzero. 
Assuming that the interactions do not break the rotation invariance,
the correlated BCS wave 
functions have only two nodal cells. 
In fact, this can hold even if the 
invariance is broken but one would need to show it for the entire
exchange path, not only for a single point.
The proof extends to $d>2$ and to other models as well, 
with details given elsewhere \cite{mitaslong}.

For the classes of fermion systems studied in this work
the two nodal cell property indeed appears as a generic feature.
That brings up an interesting question: when might this property 
not apply?
We mention just some of the
possibilities: i) additional symmetries and/or boundary
conditions can generate 
additional nodal cells;
  ii) nonlocal or very strong/singular interactions can   
  reorder the states (eg, an excited state becoming the ground state) 
or significantly change the nodes;
iii) for systems with strong correlations, large degeneracies 
at the Fermi level, or at quantum phase transitions
the properties of nodes are far from clear and require further study. 

We believe the presented analysis and proofs provide a significant 
step forward in our understanding of topologcal properties of fermionic wave
functions. The results offer an elegant and unifying framework for several
previously conjectured or numerically investigated features and 
apply to arbitrary size for both noninteracting and interacting systems. The 
presented techniques extend to other models 
and open exciting 
perspectives for studies of many-body effects which are
currently out of reach.
  

I would like to acknowledge the support by 
NSF DMR-0102668, DMR-0121361 and  EAR-0530110 grants. 

\end{document}